\documentclass[12pt]{bmc_article}

\usepackage{amsmath}

\usepackage{cite} 
\usepackage{url}  
\usepackage{ifthen}  
\usepackage{multicol}   
\usepackage[utf8]{inputenc} 
\newboolean{publ}

\newcommand{\kvec}{\boldsymbol{k}}
\newcommand{\fvec}{\boldsymbol{f}}
\newcommand{\nvec}{\boldsymbol{n}}
\newcommand{\mvec}{\boldsymbol{m}}
\newcommand{\qvec}{\boldsymbol{q}}

\usepackage{graphicx}

\newenvironment{bmcformat}{\begin{raggedright}\baselineskip20pt\sloppy\setboolean{publ}{false}}{\end{raggedright}\baselineskip20pt\sloppy}

\begin{document}
\begin{bmcformat}

\title{Gap modification of atomically thin boron nitride by phonon mediated interactions.}

\author{J.P. Hague\email{J.P.Hague@open.ac.uk}}
\address{Department of Physical Sciences, The Open University, Walton Hall, Milton Keynes, MK7 6AA, United Kingdom}



\maketitle

\begin{abstract}
A theory is presented for the modification of bandgaps in atomically
thin boron nitride (BN) by attractive interactions mediated through phonons
in a polarizable substrate, or in the BN plane. Gap equations are
solved, and gap enhancements are found to range up to 70\% for
dimensionless electron-phonon coupling $\lambda=1$, indicating that a
proportion of the measured BN bandgap may have a phonon origin.
\end{abstract}


\ifthenelse{\boolean{publ}}{\begin{multicols}{2}}{}




\section*{Introduction}

The need for bandgaps in graphene on eV scales has led to a number of
proposals, such as the use of bilayer graphene \cite{mccann2006a}
creation of nanoribbons \cite{brey2006a} and manipulation through
substrates \cite{zhou2007a,enderlein2010a}. Recently it has become
possible to manipulate atomically thin layers of boron nitride (BN)
and other materials with structure similar to graphene
\cite{novoselov2005a}. This may lead to a complimentary method of
manipulating bandgaps to make digital transistors.

In low dimensional materials, strong effective electron-electron
interactions can be induced via interaction between electrons confined
to a plane and phonons in a polarizable neighboring layer
\cite{alexandrov2002a}. Theory has shown that similar interactions
account for the transport properties of graphene on polarizable
substrates \cite{fratini2008a}, and that sandwiching graphene between
polarisable superstrates and gap opening substrates can cause gap
enhancement \cite{hague2011b}. This paper examines similar gap changes
in atomically thin BN due to interactions mediated through substrates.

\section*{Model}
\label{sec:model}

Atomically thick hexagonal BN (h-BN) has similar chemistry to
graphene: Bonding occurs through $sp_2$ hybridization, and electrons
with energies close to the chemical potential are in unhybridized
$\pi$ orbitals \cite{alem2009a}. A key difference is that the
electronic charge is not completely screened by the $sp_2$
hybridization, shifting $\pi$ orbitals by $\Delta_{\nvec}=+\Delta$ on
N sites and $-\Delta$ on B sites. This shift is the dominant cause of
a gap of order $2\Delta$. Tight binding fits to {\it ab-initio}
simulations of monolayer BN have established that the hopping,
$t=2.33$eV \cite{ribeiro2011a}, with an estimate of
$\Delta=1.96$eV$=0.84t$. Experiments indicate larger gaps: bulk h-BN
has $5.971$eV \cite{watanabe2004a}, and monolayer h-BN has a gap of
$5.56$eV \cite{song2010a} corresponding to
$\Delta=2.78$eV$=1.20t$. There is significant variation in phonon
energies, $\hbar\Omega_{\qvec}$, in h-BN \cite{serrano2007a}. LA
phonons energies range up to around 140meV at the M point, and TA
phonons to around 110meV at the K point. Optical phonon energies range
between 160meV and 200meV. Coupling, $f_{\nvec}(\mvec)$, between
electrons and phonons in either a polarisable substrate, or the BN
monolayer is possible and the corresponding Hamiltonian is,
\begin{equation}
H = -t\sum_{\langle \nvec,\nvec'\rangle\sigma}(a^{\dagger}_{\nvec\sigma}c_{\nvec'\sigma} + c^{\dagger}_{\nvec'\sigma}a_{\nvec\sigma}) - \sum_{\nvec\mvec\sigma} f_{\nvec}(\mvec)n_{\nvec\sigma}\xi_{\mvec}
+ \sum_{\mvec} \hbar\Omega (N_{\mvec}+1/2) + \sum_{\nvec\sigma}\Delta_{\nvec}n_{\nvec\sigma}.
\end{equation}
The Hamiltonian terms are shown schematically in Fig. \ref{fig:summary}
(left). $a^{\dagger}_{\nvec\sigma}$ creates electrons of spin $\sigma$ on B sites and $c^{\dagger}_{\nvec'\sigma}$ on N sites. Vectors
$\nvec$ are to atoms in the monolayer, and $\mvec$ to atoms in the
substrate. $N_{\mvec}$ is the number operator for phonons. The
Hamiltonian is also approximately valid for interactions in the plane,
and Fig. \ref{fig:summary} (right) shows the forces on ions from an
increase in electron density at a B site. The largest forces are on
the near-neighbor sites, so the effective interaction is mainly
site diagonal (electrons on A sites self-interact through phonons on B
sites and vice versa). The diagram indicates that the strongest
interaction is between electrons and optical phonon modes.

For simplicity, I use the Holstein electron-phonon interaction,
$f_{\nvec}(\mvec)\propto\delta_{\nvec,\nvec'}$, which qualitatively
captures the physics. There may be quantitative changes to the results
for longer range Fr\"ohlich interactions and from modulation of the
electron-phonon interaction due to incommensurability of the substrate, which
was estimated at around $\pm 8\%$ of the average value
\cite{hague2011b}.


\section*{Gap equations and results}
\label{sec:low}

Low order perturbation theory is applicable for low phonon frequency
and weak coupling. I derive a set of gap equations by symmetrizing the
self energy,
\begin{equation*}
\mathbf{\Sigma}(i\omega_n) \approx \left(
\begin{array}{cc}
i\hbar\omega_n(1-Z^{A}_n)+\bar{\Delta}^{A}_n & 0 \\
0 & i\hbar\omega_n(1-Z^{B}_n)-\bar{\Delta}^{B}_n
\end{array}
\right).
\end{equation*}
The local approximation used here is a good starting point here
because the modulated potential $\Delta$ is large, and electrons are
well localised. Off diagonal terms do not feature in the lowest order
perturbation theory for the Holstein model since the interaction is
site diagonal. $Z_n$ is the quasi-particle weight and $\bar{\Delta}_n$
is the gap function. For bosonic quantities, $\hbar\omega_s = 2\pi k_B
T s$ and for fermions $\hbar\omega_n = 2\pi k_B T (n + 1/2)$.
$T$ is the temperature and $n$ and $s$ are integers.

The full Green function can be established using Dyson's equation $\mathbf{G}^{-1}(\kvec,i\omega_n) = \mathbf{G}_0^{-1}(\kvec,i\omega_n) - \mathbf{\Sigma}(i\omega_n)$, leading to,
\begin{equation}
\mathbf{G}^{-1}(\kvec,i\omega_n) = \left(
\begin{array}{cc}
Z^{A}_ni\hbar\omega_n - \Delta - \bar{\Delta}^{A}_n & \phi_{\kvec}^{*} \\
\phi_{\kvec} & Z^{B}_ni\hbar\omega_n + \Delta+\bar{\Delta}^{B}_n
\end{array}
\right).
\end{equation}

Substituting the expression for the Green function into the lowest order contribution to the self energy,
\begin{equation}
\Sigma_{ii}(\kvec,i\omega_n) = -k_BTt\lambda_i\sum_{s}\int \frac{{\rm d}^2\qvec}{V_{BZ}} G_{ii}(\kvec-\qvec,i\omega_{n-s})(d^{(ii)}_0(\qvec,\omega_s)-2d^{(ii)}_0(0,0)).
\end{equation}
Here, the phonon propagator,
$d^{(ij)}_0(\qvec,\omega_s)=\delta_{ij}\Omega^2/(\Omega^2+\omega_s^2)$,
hence there are no off diagonal elements of the lowest order self
energy. The use of a single averaged $\Omega$ and $\lambda$ here is
consistent with a mean-field approximation. At half-filling, it is
reasonable to assume that $\lambda_A=\lambda_B$ within the scope of
the model, so that $\Delta^{A}=\Delta^{B}$ and $Z^{A}=Z^{B}$. This
leads to the gap equations,
\begin{equation}
\bar{\Delta}_n = - t\lambda k_B T \sum_s \int {\rm d}\epsilon \frac{D(\epsilon)\, \Delta'_{n-s}(d_0(i\omega_s)-2)}{\hbar^2 \omega_{n-s}^2 Z^2_{n-s}+\Delta'^{2}_{n-s}+\epsilon^2},
\end{equation}
\begin{equation}
Z_n = 1-\frac{t\lambda k_B T}{\hbar\omega_n} \sum_s \int {\rm d}\epsilon \frac{D(\epsilon)\,\omega_{n-s}Z_{n-s}d_0(i\omega_s)}{\hbar^2\omega_{n-s}^{2}Z^2_{n-s}+\Delta'^2_{n-s}+\epsilon^2},
\label{eqn:qpweight1}
\end{equation}
where the full gap is $\Delta'_n = \bar{\Delta}_n+\Delta$.  The
density of states for a tight binding hexagonal lattice in the absence
of a gap, $D(\epsilon)$, has the form given in
Ref. \cite{castroneto2009a}. The equations may be solved self
consistently by performing a truncated sum on Matsubara
frequencies.



Gap and quasi-particle weight functions only have a weak Matsubara
frequency dependence ($<0.3\%$ for $\lambda=1$,
$k_{B}T=\hbar\Omega=0.01t$).  The local gap enhancement factors
$\Delta'/\Delta$, are shown in Fig. \ref{fig:eliashberg} (left) for
various $\lambda$, showing a modest increase of $\sim 70\%$ for
$\lambda=1$. The enhancement factor increases slightly with decreasing
$\Delta$ but is essentially unchanged by modifications to phonon
frequency and temperature for the parameter values used here. I also
calculate the temperature dependence of the gap,
Fig. \ref{fig:eliashberg} (right). For very large temperatures, where
$k_BT$ approaches $\Delta$, there is a drop in the gap size. For $T<
0.3t\sim$8000 K, this levels off, and the gap becomes relatively
constant.


\section*{Summary and conclusions}
\label{sec:summary}

I have presented a theory for the modification of BN band-gaps by
interaction with phonons. It is of interest to make comparison between
the bandgaps of bulk h-BN, nanotubes, monolayer h-BN and the theory
presented here. Measured bandgaps of bulk h-BN are of between 5.8eV
\cite{zunger1976a} and 5.971eV \cite{watanabe2004a}, indicating that
interaction between layers increases the bandgap, consistent with the
theory here. The bulk gap is also higher than that for nanotubes (5eV) \cite{terauchi1998a}. On the other hand, Song {\it et al.} \cite{song2010a}
claim that the gap is reduced as BN thickness increases. The above
discussion is presented with the caveat that the theory requires that
hopping between the substrate and the BN monolayer is
small. Interlayer hopping will affect the bandwidth and bandgap, and
direct Coulomb interaction with strongly ionic substrates could also
affect the band structure if the charge density at the surface of the
substrate varies dramatically.

It is also of interest to estimate the magnitude of the bandgap
modification due to electron-phonon interaction in isolated monolayers
of BN. {\it Ab-initio} calculations have attempted to quantify the
magnitude of the interaction between electrons and acoustic phonons
for small momentum excitations \cite{bruzzone2011a}. Extrapolating the
interaction, and taking a mean-field average (assuming mean momentum
magnitude of $4\pi/9a$), the electron-phonon coupling can be estimated as
$\lambda=(4\pi/9)^2 E_1^2/2 a^2 t \bar{M} \Omega^2$, taking
$E_{1}=3.66$eV from Ref. \cite{bruzzone2011a}, $\bar{M}\approx 12.5$
amu, $a=2.5\AA$. The mean energy of longitudinal acoustic phonons lies in the range $50-75$meV, giving a range of $\lambda=0.05-0.12$,
so the contribution of phonons to the bandgap is estimated as
$3-7$\%. I would expect BN to have stronger interaction with optical
phonons, since the pattern of distortions around an electronic defect
is consistent with optical modes (see Fig. 1).

The BN gap is too wide for digital applications. Recently, it has become possible to manufacture silicene, an
atomically thick layer of silicon with similar properties to graphene
\cite{padova2010a}, so it may be possible to make GaAs or AlP
analogues to BN. Smaller gaps could be available from those materials,
which might be used to create tunable bandgaps for atomically thick
transistors.

\section*{Acknowledgments}

I acknowledge EPSRC grant EP/H015655/1 for funding and useful
discussions with A. Ilie and A. Davenport.


\newpage
{\ifthenelse{\boolean{publ}}{\footnotesize}{\small}
 \bibliographystyle{bmc_article}  
  \bibliography{atom_boron_nitride_rev} }     


\ifthenelse{\boolean{publ}}{\end{multicols}}{}

\begin{figure}
\includegraphics[width = 65mm]{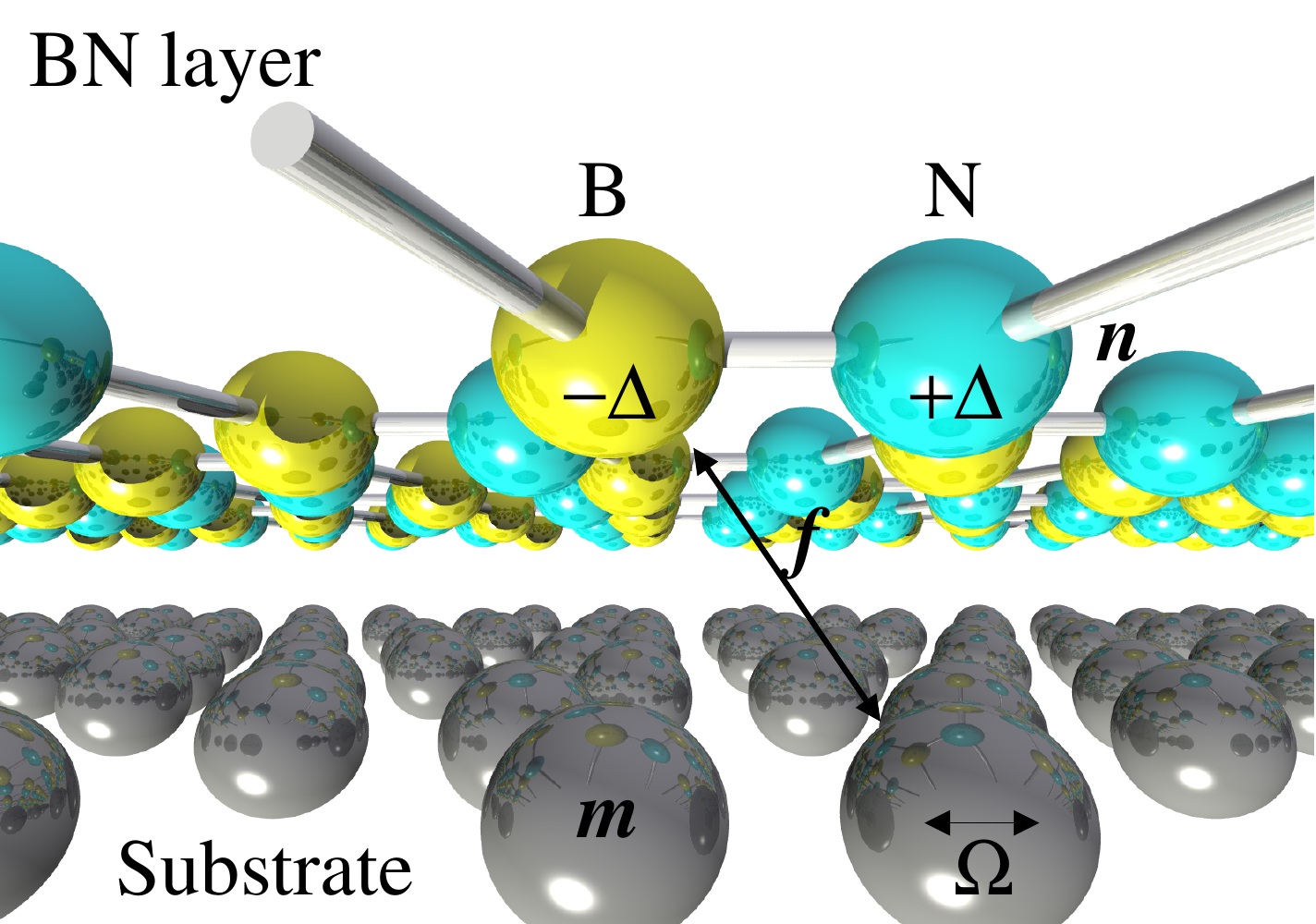}
\includegraphics[width = 45mm]{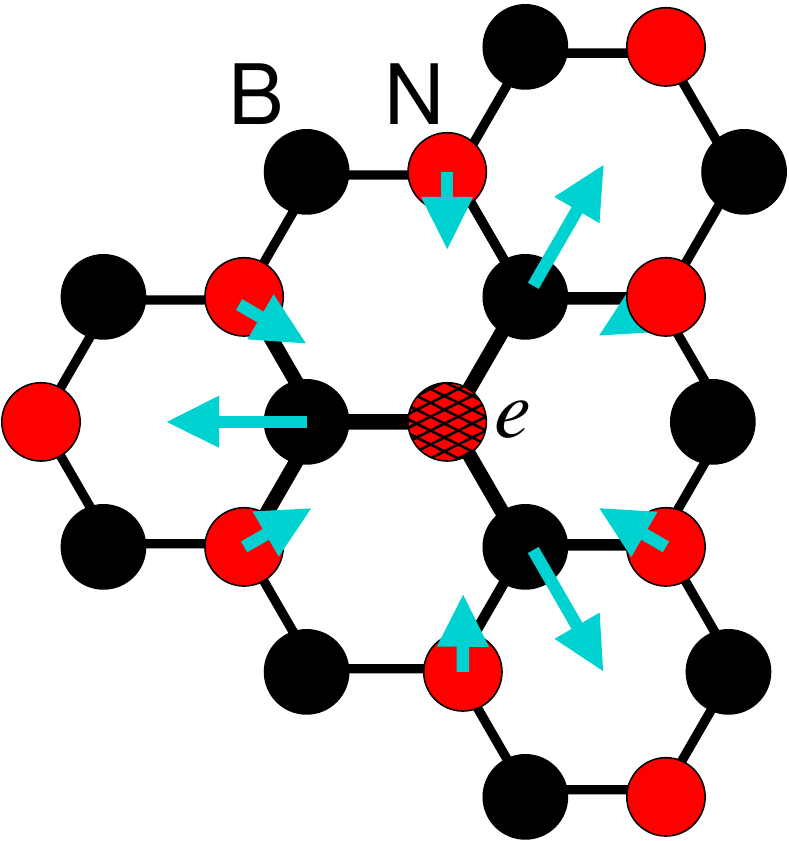}
\caption{(left) BN-substrate system annotated with
  interactions. Electron-phonon interactions between the BN layer and
  substrate are poorly screened, and large interactions of strength
  $f_{\nvec}(\mvec)$ are possible. Ions in the substrate oscillate
  with frequency $\Omega$. N sites have energy $+\Delta$ and B sites
  $-\Delta$, opening a gap.  The attractive phonon mediated electronic
  interaction $\fvec$ binds electrons onto the same site, effectively
  enhancing the gap. (right) Interactions
  in a monolayer of BN. \label{fig:summary}}
\end{figure}

\begin{figure}
\includegraphics[width = 75mm]{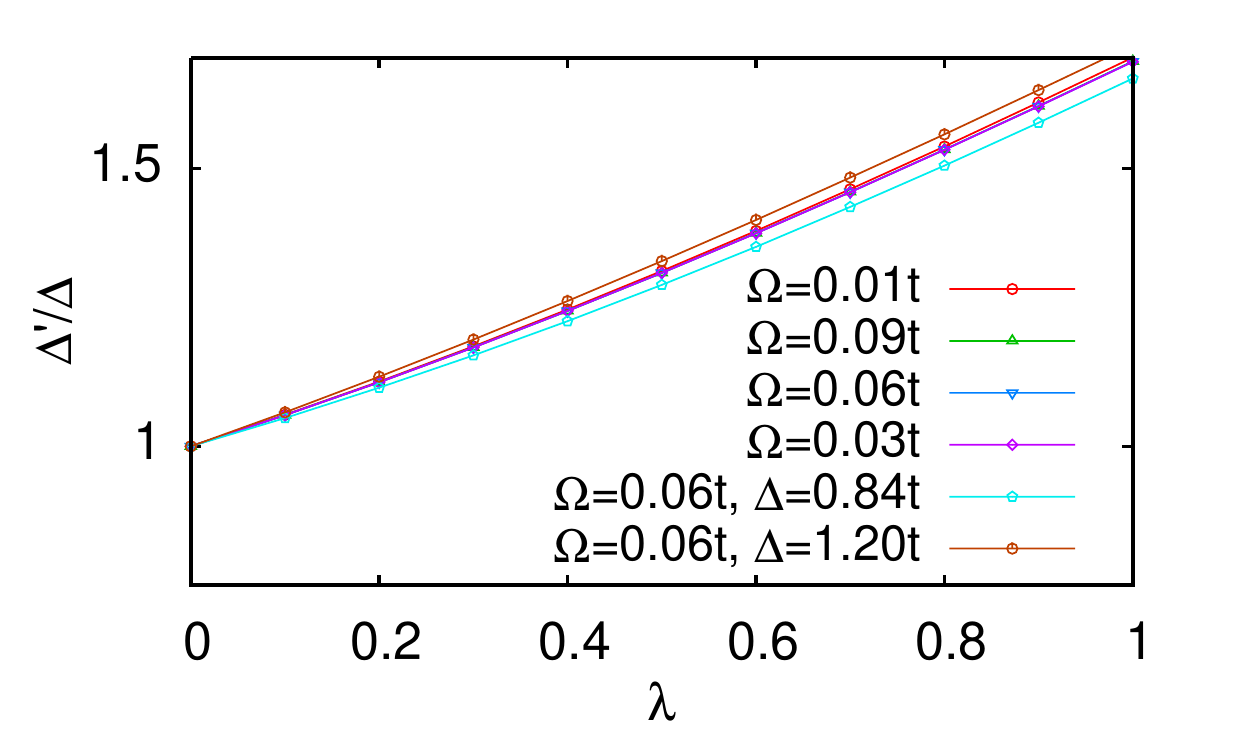}
\includegraphics[width = 75mm]{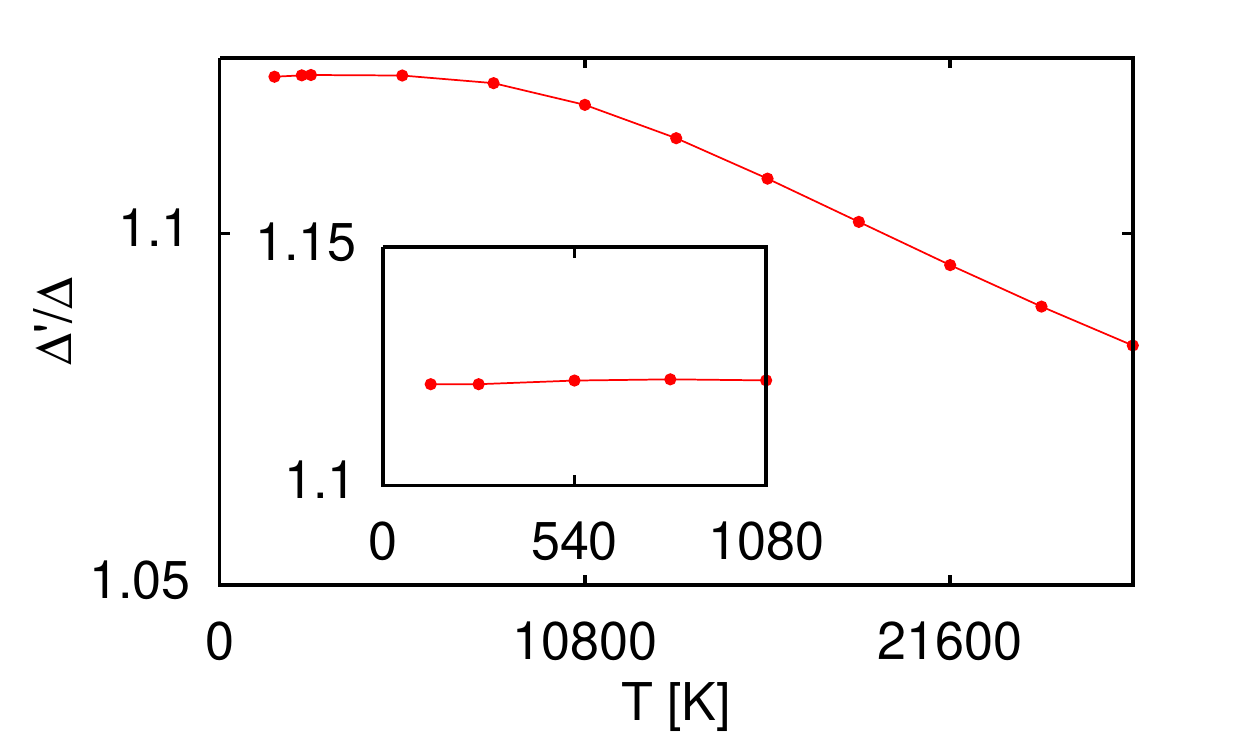}
\caption{(left) Modification of the BN bandgap. The gap enhancement depends mainly on $\lambda$, is weakly dependent on $\Delta$ and shows almost no change with $\Omega$. Calculations are made for $\Delta=t$ corresponding
  to a BN gap of $2\Delta = 4.66$eV, $\Delta=1.20t$ ($2\Delta = 5.6$eV), and $\Delta=0.84t$ ($2\Delta=3.92$eV, the tight binding fit from Ref. \cite{ribeiro2011a}). $t=2.33$eV,
  $\hbar\Omega=0.01t=23$meV, $\hbar\Omega=0.03t=70$meV, $\hbar\Omega=0.06t=140$meV, $\hbar\Omega=0.09t=210$meV, covering the full range of phonon frequencies in Ref. \cite{serrano2007a}. $k_B T = 0.01t$ ($T=268$K) and $\lambda\le 1$ (right) Variation of the gap with temperature, $\hbar\Omega=0.05t=117$meV and $\lambda=0.2$. There is a weak temperature dependence, due to the large $\Delta$, consistent with the measurements in Ref. \cite{zunger1976a}, with the gap starting to close only for extremely high temperatures $T>8000K$, presumably above the melting point of the material.}
\label{fig:eliashberg}
\end{figure}


\end{bmcformat}
\end{document}